\documentclass[a4,12pt,epsf]{article}
\usepackage{amssymb}
\usepackage{amsmath}
\usepackage{hyperref}
\usepackage{multirow}
\usepackage{graphicx}
\usepackage{pstricks}
\usepackage{color}
\usepackage{setspace}

\begin{document}
%%%%% title %%%%%
\begin{center}{\Large \bf
Generation of particle number asymmetry in expanding universe\footnote{contributed talk at 1st workshop on Phenomenology for Particle and Anti-Particle 2018 (PPAP 2018)}
%\vspace{6pt}%\\
%short talk
}
\end{center}
%%%%% author(s) %%%%%
\begin{center}
Apriadi Salim Adam%$^a$
\footnote{apriadiadam@hiroshima-u.ac.jp}
and
Takuya Morozumi%$^a$
\footnote{morozumi@hiroshima-u.ac.jp}
\vspace{6pt}\\
%%%%% address(es) %%%%%
%$^a$
{\it
Graduate School of Science Hiroshima University,
1-3-1 Kagamiyama, Higashi-Hiroshima, 739-8526, Japan\\
Core of Research for Energetic Universe, Hiroshima University, Higashi-Hiroshima,
739-8526, Japan
}
\vspace{6pt}\\
Keiko I. Nagao%$^a$
\footnote{nagao@dap.ous.ac.jp}
\vspace{6pt}\\
%%%%% address(es) %%%%%
%$^a$
{\it
	Okayama University of Science, Ridaicho, Kita-ku, Okayama, 700-0005, Japan
}
\vspace{6pt}\\
Hiroyuki Takata%$^a$
\footnote{takata@tspu.edu.ru}
\vspace{6pt}\\
{\it
	Tomsk State Pedagogical University, Tomsk, 634061, Russia
}
% Anoterh Address
%{
%\\
%$^b$
%{\it  another address
%Yukawa Institute for Theoretical Physics, YITP,
%Kyoto 606-8502, Japan
%}
\end{center}
%%%%% abstruct %%%%%
\begin{abstract}
We study the creation and time evolution of particle number asymmetry with nonequilibrium quantum field theory. We introduce a model of a neutral scalar and a complex scalar and it has CP violating and particle number violating features. Starting with an initial condition specified by density operator, we show how particle number asymmetry can be generated through interaction. We investigate the time evolution of particle number asymmetry of the universe using the perturbation method. 
\end{abstract}
%%%%%%%%%%%%%%%%%%%%
%\section{Background}
\section{Introduction}

What still not yet fully understood from the standard model (SM) point of view is why there is more baryon than anti-baryon in the universe \cite{Kne2004}. In order to address this issue, a number of studies have been proposed \cite{fuk86}. In present work, we study a model which generates particle number asymmetry through interactions and develop a formulation which is applicable to various types of expanding universe \cite{mor2017,apri2017}. In addition, we also compute the time evolution of current asymmetry by using quantum field theory with the density operator.

\section{The model}

%\subsection{Introduction}
The model consists of a neutral scalar, $N$, and a complex scalar, $\phi$. The action $S$ is given by \cite{mor2017,apri2017},
\begin{equation}
\label{Lageq}
\begin{aligned}
S =& \int d^{4}x \sqrt{-g} \left(\mathcal{L}_{\text{free}}+\mathcal{L}_{\text{int}} + \xi (R - 2 \Lambda)\right),  \\
\mathcal{L}_{\text{free}} =& g^{\mu\nu} \nabla_{\mu} \phi^{\dagger}
\nabla_{\nu} \phi - m_{\phi}^{2} | \phi |^{2} + \frac{1}{2} \nabla_{\mu} N \nabla^{\mu} N  - \frac{M_N^2}{2} N^{2}+ \frac{B^2}{2}  (\phi^2 + \phi^{\dagger 2})    \\
&+ \left(\frac{\alpha_2}{2} \phi^2 + h.c.\right)R+ \alpha_{3} |\phi|^{2}R , \\
\mathcal{L}_{\text{int}} =& 
A\phi^2 N + A^{\ast} \phi^{\dagger 2} N + A_{0}|\phi|^{2}N ,
\end{aligned}
\end{equation}
where $A$ is the interaction coupling of the vertex, $B$ is the parameter which gives the mass difference of the fields and $\alpha_{2}$ is the matter-curvature coupling. The metric $g_{\mu\nu}$ is given by the type of Friedmann Robertson Walker with scale factor $a(x^{0})$,
\begin{align}
g_{\mu \nu} = (1, -a^2(x^0), -a^2(x^0), -a^2(x^0)),
\end{align}
while $R$ is the Riemann curvature which has form $R = 6 \left[\left(\frac{\ddot{a}}{a}\right)+ \left(\frac{\dot{a}}{a} \right)^2\right]$ and $H(x^{0}) = \frac{\dot{a}}{a}$. With the Lagrangian in Eq.\eqref{Lageq}, we study the particle number asymmetry produced by the soft-breaking terms of U(1) symmetry whose coefficients are denoted by $A$ and $B^2$. Noticing that, with this Lagrangian, one can also derive the Einstein equations for the scale factor coupled with scalar particles. They are given as,
\begin{align}
T_{\mu \nu}  = & \partial_{\mu} \phi_i \partial_{\nu} \phi_i - g_{\mu \nu}
\left( \frac{1}{2} g^{\alpha \beta} \partial_{\alpha} \phi_i
\partial_{\beta} \phi_i - \frac{1}{2} m_i^2 \phi_i^2 + \mathcal{}
\frac{1}{3} A_{ij k} \phi_i \phi_j \phi_k \right) \\
	 - 3 (1 - 8 \pi G \beta_i \phi_i^2)  \left(
	\frac{\dot{a}}{a} \right)^2 + \Lambda = & - 8 \pi G T_{00}\ (\text{$00$ component})\\
	 (1 - 8 \pi G \beta_i \phi_i^2) (2 a
	\ddot{a} + \dot{a}^2) - a^2 \Lambda = & - 8 \pi G T_{i i}\ (\text{$ii$ component}) \\
	 0 = & - 8 \pi G T_{\mu \nu (\neq \mu)} \ (\text{off diagonal component})
\end{align}
However, a full discussion for solving them lies beyond the scope of this study. At present, we work for the case that the time dependence of the scale factor is given.

\section{The current expectation value and its time evolution}

The particle number is related to U$(1)$ transformation of the complex scalar field, namely, $\phi(x)\rightarrow \phi (x) e^{i\theta}$. It is U$(1)$ charge represented by particle number operator $N$ \cite{AffDin85},
%U$(1)$ transformation is related to 
\begin{align}
N(x^{0}) =& \int \sqrt{-g(x)} j_{0}(x) d^{3}{\bf x} \\
j_{\mu}(x) =& i (\phi^{\dagger} \partial_{\mu} \phi -\partial_{\mu} \phi^{\dagger} \phi).
\end{align}
As for the initial condition of the state, it is given by density operator, namely,
\begin{equation}\label{key}
\rho(t_0) = \frac{e^{- H_0/T}}{\mathrm{tr} e^{- H_0/T}},
\end{equation}
where $H_0$ is a Hamiltonian includes the linear term of the fields and $T$ denotes the temperature.

It is convenient to write the scalar particles in terms of real fields by using the following relations, $\phi \equiv\frac{{\phi_1} + {i \phi_2}}{\sqrt{2}}$ and $\phi_3 \equiv N$. In terms of real fields, the current expectation value written with initial density operator has form,
\begin{align}
\langle j_{\mu}(x)\rangle &=\mathrm{tr} (j_{\mu} (x) \rho(t_0)), \nonumber \\
&= \text{Re.} \left[   \left( \frac{\partial}{\partial x^{\mu}} -
\frac{\partial}{\partial y^{\mu}} \right) G_{12}^{} (x, y) \big|_{y
	\rightarrow x}  + \bar{\phi}^{\ast}_2 (x)\overset{\leftrightarrow}{\partial_{\mu}}  \bar{\phi}_1 (x)  \right] , \label{jmu}
\end{align}
where $\bar{\phi}$ denotes the mean field with a relation, $\phi=\bar{\phi}+\varphi$. Both Green's function and field are obtained from 2PI effective action $\Gamma_{2}$ \cite{ramsey97},
\begin{align}
\Gamma_2 [G, \bar{\phi}, g] = S[\bar{\phi}, g] + \frac{i}{2}
\text{TrLn}\ G^{- 1} + \Gamma_Q - \frac{i}{2} \text{Tr}\ {\bf 1} + \frac{1}{2}  \int d^4 x \int d^4 y
\frac{\delta^2 S[\bar{\phi}, g]}{\delta \bar{\phi}_i^a (x) \delta \bar{\phi}_j^b (y)}  G^{ab}_{ij} (x, y),
\end{align}
where $\Gamma_Q$ is the lowest order of 2PI diagram. Their equations of motion are derived by taking derivative $\Gamma_{2}$ with respect to Green's function and field, respectively. 

Before closing this section, let us write down the time evolution of the current asymmetry. First, we consider the solution of Green's function and field up to the first order of interaction coupling $A$ and up to the linear order of the Hubble constant $H(t_0)$. In this case, we focus on the case that the initial expectation value of the field is $(\bar{\phi}_1,\bar{\phi}_2,\bar{\phi}_3)=(0,0,v_3)$. Then, the contribution to the current asymmetry is only given by the first term of Eq.\eqref{jmu},
\begin{align}
\langle j_0 (x^0) \rangle_{O(A)} = & \int
	\frac{d^3 k}{(2 \pi)^3}   \left( \frac{\partial}{\partial x^0} -
	\frac{\partial}{\partial y^0} \right) [\text{Re.} \hat{G}_{12}^{O(A)} (x^0, y^0, {\bf k})] \big|_{y^{0} \rightarrow x^0},  \nonumber \\
	=& \langle j_0 (x^0) \rangle_{  \text{1st} } + 	\langle j_0 (x^0) \rangle_{ \text{2nd} } ,
\end{align}
where $\langle j_0 (x^0) \rangle_{  \text{1st} }$ and $\langle j_0 (x^0) \rangle_{ \text{2nd} }$ are given by,
{\small 
	\begin{align}
	\langle j_0 (x^0) \rangle_{  \text{1st} }  =& \frac{2
		v_{3} A_{1 2 3}}{a_{t_0}^3} \int \frac{d^3 
		{\bf k}}{(2 \pi)^3} \int_{t_0}^{x^0} \left\{ 1 - 3 (x^0 - t_0) H (t_0)  -
		\frac{3}{2} (t - t_0) H (t_0) \right\} \nonumber\\
	& \times \left[ \left\{ \frac{(- \bar{K}^{(0)\prime}_{3, t t_0,
			{\bf 0}} )}{2 \omega_{2, {\bf k}} (t_0) } \coth \frac{\beta
		\omega_{2, {\bf k}} (t_0)}{2} \left[ \left( \bar{K}^{(0)\prime}_{2,^{}
		x^0 t_0, {\bf k}} \overset{\leftrightarrow}{\partial\ \dot{}}
	\bar{K}^{(0)}_{1, x^0 t, {\bf k}} \right) \bar{K}^{(0)\prime}_{2, t
		t_0, {\bf k}} \right. \right. \right. \nonumber\\
	&   \left. \left. \left. + \omega^2_{2, {\bf k}} (t_0) \left(
	\bar{K}^{(0)}_{2, x^0 t_0, {\bf k}} \overset{\leftrightarrow}{\partial\
		\dot{}} \bar{K}^{(0)}_{1, x^0 t, {\bf k}} \right) \bar{K}^{(0)}_{2, t
		t_0, {\bf k}} \right] \right\} - \{ 1 \leftrightarrow 2\ \text{for lower indices} \} \right] d t, \label{j01}\\
	%\end{align}
	%\begin{align}
	\langle j_0 (x^0) \rangle_{ \text{2nd} } =&   \frac{2
		v_{3} A_{1 2 3}}{a_{t_0}^3} \int \frac{d^3 
		{\bf k}}{(2 \pi)^3} \int_{t_0}^{x^0}  \left[ \left\{
	\frac{(- \bar{K}^{(0)\prime}_{3, t t_0, {\bf 0}} )}{2 \omega_{2,
			{\bf k}} (t_0) } \coth \frac{\beta \omega_{2, {\bf k}} (t_0)}{2}
	\right. \right. \nonumber\\
	& \times \left[ \left( \bar{K}^{(0)\prime}_{2,^{} x^0 t_0, {\bf k}}
	\overset{\leftrightarrow}{\partial\ \dot{}} \bar{K}^{(0)}_{1, x^0 t,
		{\bf k}} \right) \bar{K}^{(1)\prime}_{2, t t_0, {\bf k}} \right.
	+ \left( \bar{K}^{(1)\prime}_{2,^{} x^0 t_0, {\bf k}}
	\overset{\leftrightarrow}{\partial\ \dot{}} \bar{K}^{(0)}_{1, x^0 t,
		{\bf k}} + \bar{K}^{(0)\prime}_{2,^{} x^0 t_0, {\bf k}}
	\overset{\leftrightarrow}{\partial\ \dot{}} \bar{K}^{(1)}_{1, x^0 t,
		{\bf k}} \right) \bar{K}^{(0)\prime}_{2, t t_0, {\bf k}}
	\nonumber\\
	& + \omega^2_{2, {\bf k}} (t_0) \left[ \left( \bar{K}^{(0)}_{2, x^0
		t_0, {\bf k}} \overset{\leftrightarrow}{\partial\ \dot{}}
	\bar{K}^{(0)}_{1, x^0 t, {\bf k}} \right) \bar{K}^{(1)}_{2, t t_0,
		{\bf k}} \right. \nonumber\\
	& \left. \left. \left.\left. + \left( \bar{K}^{(1)}_{2, x^0 t_0, {\bf k}}
	\overset{\leftrightarrow}{\partial\ \dot{}} \bar{K}^{(0)}_{1, x^0 t,
		{\bf k}} + \bar{K}^{(0)}_{2, x^0 t_0, {\bf k}}
	\overset{\leftrightarrow}{\partial\ \dot{}} \bar{K}^{(1)}_{1, x^0 t,
		{\bf k}} \right) \bar{K}^{(0)}_{2, t t_0, {\bf k}} \right] \right]
	\right\}  - \{ 1 \leftrightarrow 2\ \text{for lower indices} \}  \right] d t.  \label{j02}
	\end{align}
}
where $\bar{K}_i [x^0, y^0]$ and its derivative are given by,
{\small 
\begin{align}
\bar{K}^{(0)}_i [x^0, y^0] = & \frac{\sin [\omega_{i, {\bf  k}} (x^0 - y^0)]}{\omega_{i, {\bf  k}}},\quad \omega_{i, {\bf k}} =\sqrt{\frac{{\bf k}^2}{a(t_0)^{2}}+\tilde{m}_i^2},\nonumber  \\
\tilde{m}_1^2 =& m_{\phi}^2 - B^2,\quad \tilde{m}_2^2 = m_{\phi}^2 + B^2,\quad \tilde{m}_3^2 = m_N^2, \label{k0}\\
\bar{K}^{(1)}_i[x^0, y^0] =&  \frac{H (t_0){\bf  k}^2 (x^0 + y^0 - 2 t_0)}{2\omega_{i, {\bf  k}}^2 a (t_0)^2}
 \left( \frac{\sin [\omega_{i, {\bf  k}} (x^0 -
	y^0)]}{\omega_{i, {\bf  k}}} - (x^0 - y^0) \cos [\omega_{i, {\bf  k}}
(x^0 - y^0)] \right), \label{k1} \\
\bar{K}'_i [x^0, y^0]:=& \frac{\partial \bar{K}_{i}[x^0,y^0]}{\partial y^0},\quad \dot{\bar{K}}_i [x^0, y^0]:=\frac{\partial \bar{K}_{i}[x^0,y^0]}{\partial x^0},\quad \dot{\bar{K}}'_i [x^0, y^0]:=\frac{\partial^2 \bar{K}_{i}[x^0,y^0]}{\partial x^0 \partial y^0}, 
\end{align}
}
From Eqs.\eqref{j01} and \eqref{j02}, below we remark several findings:
\begin{enumerate}
	\item In the first line of Eq.\eqref{j01}, the first, second and third terms in wave parentheses correspond to the contribution of constant scale factor to the current asymmetry, dilution effect and freezing interaction effect, respectively. 
	\item Eq.\eqref{j02} corresponds to the redshift effect for the contribution to the current asymmetry.
\end{enumerate}
\begin{table}
	\begin{center}
		\caption{The classification of $o(H(t_0))$ contributions to the current asymmetry \cite{apri2017}.}
		\begin{tabular}{|l|p{8cm}|} \hline \hline 
			The effect & The origin \\ \hline \hline
			Dilution & The increase of volume of the universe due to expansion, $\frac{1}{a(x^0)^{3}}-\frac{1}{a(t_0)^{3}}$\\ \hline 
			Freezing interaction & The decrease of the strength of the cubic interaction as $\left\lbrace  \left( \frac{a(t_0)}{a(x^0)}\right) ^{3/2}-1\right\rbrace  A_{123} $.\\ \hline
			Redshift & The effective energy of particle, $\frac{{\bf k}^{2}}{a (x^0)^2} + \tilde{m}_i^2 $. \\ \hline \hline 
		\end{tabular}
		\label{tb3}
	\end{center}
\end{table} 
The above four types of the contribution to the current asymmetry are explained as follows. The constant scale factor which is the zeroth order of $H(t_0)$ is the leading contribution. The rests which are the first order term contribute according to their origins and we summarize them in Table \ref{tb3}. In the next section, we will study numerically the time evolution of the current asymmetry.

\section{Numerical results}
\begin{figure}[h]
	\centering
	\begin{tabular}{c c}
		\includegraphics[width=.47\textwidth]{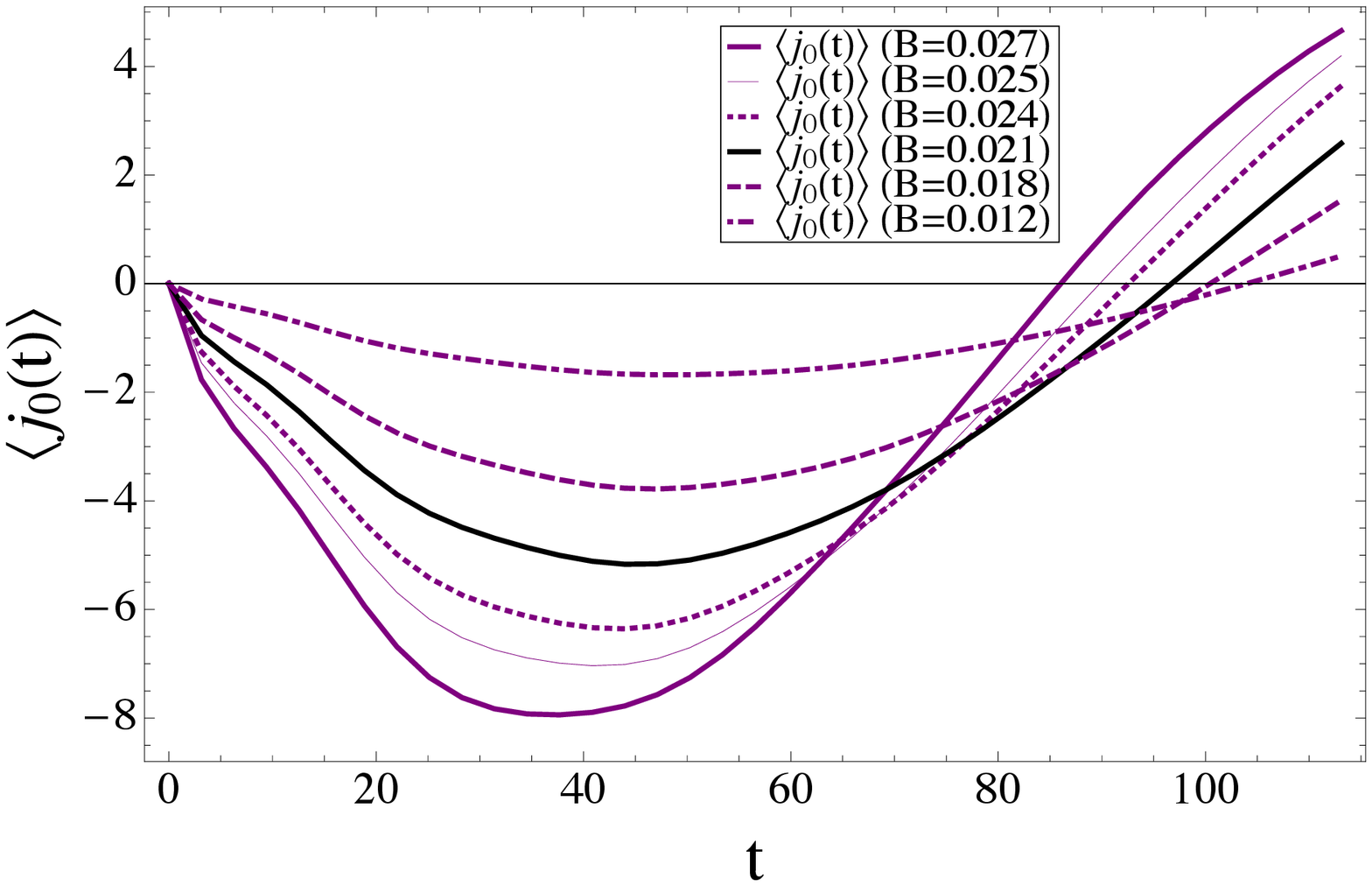}
		&
		\includegraphics[width=.47\textwidth]{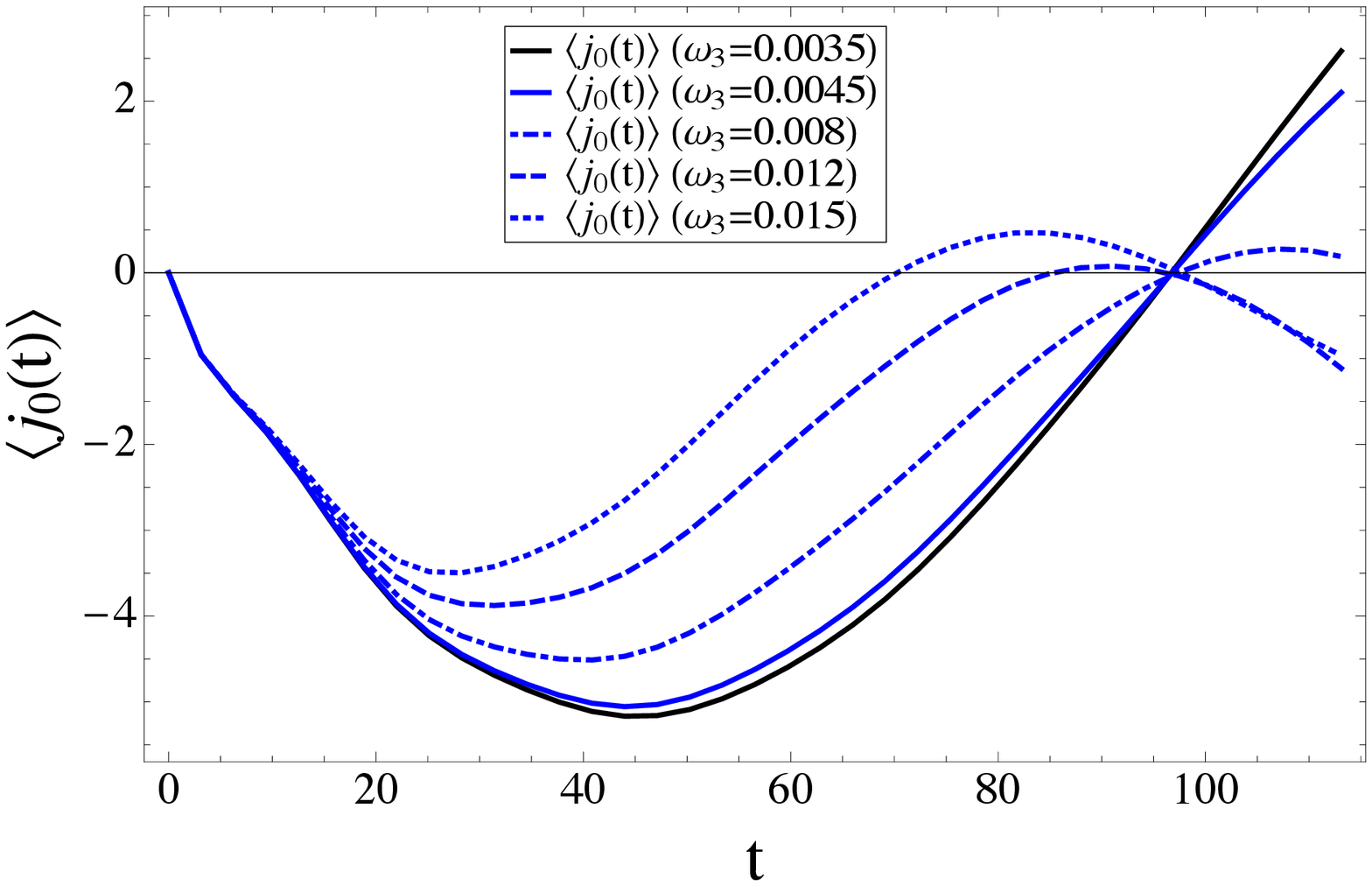}
	\end{tabular}
	\caption{\label{fig01} $B$ dependence (left) and $\omega_{3, {\bf 0}}$ dependence (right) for the time evolution of PNA are displayed. In both figures, the horizontal axis is the dimensionless time defined as $t = \omega_{3, {\bf 0}}^r (x^0 -t_0)$ and we choose $\omega_{3, {\bf 0}}^r = 0.35$ as a reference angular frequency. In the left figure, we fix a set of parameters as ($\tilde{m}_2, T, H_{t_0}, \omega_{3, {\bf 0}}$)=($0.05,100,10^{- 3},0.0035$) for all the lines. In the right figure, we use a set of parameters as ($\tilde{m}_1, \tilde{m}_2, B,T, H_{t_0}$)=($0.04,0.05,0.021,100,10^{- 3}$) for all the lines \cite{apri2017}. }
\end{figure}

In the left side of Fig.\ref{fig01}, we show the parameter $B$ dependence. Both of the amplitude and the period of the oscillation change when we alter the parameter $B$. As it increases, the amplitude becomes larger and its period becomes shorter. The right side of Fig.\ref{fig01} shows the dependence of the PNA on $\omega_{3, {\bf 0}}$. As shown in the black, blue and dot-dashed blue lines, the position of the first node does not change when $\omega_{3, {\bf 0}}$ takes its value within the difference of $\tilde{m}_1$ and $\tilde{m}_2$. However, the amplitude of oscillation gradually decreases as $\omega_{3, {\bf 0}}$ increases up to the mass difference. The interesting findings were observed when $\omega_{3, {\bf 0}}$ becomes larger than the mass difference. As $\omega_{3, {\bf 0}}$ becomes larger, the amplitude decreases and the new node is formed at once. The dashed and dotted blue lines show this behavior.    

\section{Conclusion}
We have studied an interacting model in which particle number asymmetry is generated through interactions of scalar fields. The current for the particle and anti-particle asymmetry is given up to the first order of $A$ and linear $H(t_0)$. Time evolution of the particle number asymmetry and its parameter dependence is investigated.

%%%%% acknowledgement %%%%%
\vspace*{12pt}
\noindent
\\
{\bf Acknowledgement}
\vspace*{6pt}
\noindent
This work is supported by JSPS KAKENHI Grant Number JP17K05418 (T.M.) and supported in part by JSPS Grant-in-Aid for Scientific Research for Young Scientists (B) 26800151 (K.I.N.). 
%%%%% references %%%%%

\end{document}